\documentclass[a4paper]{article}
\usepackage[english]{babel}
\usepackage[round]{natbib}
\usepackage{authblk,bbm,bm}
\usepackage{amsmath,setspace,geometry,lineno,amssymb}
\usepackage{graphicx,url}
\usepackage[colorinlistoftodos]{todonotes}
\usepackage{algorithmic}
\usepackage[boxed]{algorithm2e}
\usepackage{listings}
\usepackage[boxed]{algorithm2e}
\usepackage{listings}
\usepackage{bm,bbm}
\usepackage{amssymb}
\usepackage{graphicx,xcolor,geometry}
\begin{document}
\begin{spacing}{1.2}

\title{A multivariate mixed hidden Markov model to analyze blue whale diving behaviour during controlled sound exposures}
  
\author[1,2]{Stacy L.\ DeRuiter}
\author[2,3]{Roland Langrock}
\author[2]{Tomas Skirbutas}
\author[4]{Jeremy A.\ Goldbogen}
\author[5]{John Chalambokidis}
\author[6,7]{Ari S.\ Friedlaender}
\author[7]{Brandon L.\ Southall}
\affil[1]{Calvin College, USA}
\affil[2]{University of St Andrews, UK}
\affil[3]{Bielefeld University, Germany}
\affil[4]{Stanford University, USA}
\affil[5]{Cascadia Research Collective, USA}
\affil[6]{Oregon State University, USA}
\affil[7]{SEA, Inc., USA}

\date{}
\maketitle

\begin{abstract} 
Characterization of multivariate time series of behaviour data from animal-borne sensors is challenging.  Biologists require methods to objectively quantify baseline behaviour, then assess behaviour changes in response to environmental stimuli. Here, we apply hidden Markov models (HMMs) to characterize blue whale movement and diving behaviour, identifying latent states corresponding to three main underlying behaviour states: shallow feeding, travelling, and deep feeding. The model formulation accounts for inter-whale differences via a computationally efficient discrete random effect, and measures potential effects of experimental acoustic disturbance on between-state transition probabilities. We identify clear differences in blue whale disturbance response depending on the behavioural context during exposure, with whales less likely to initiate deep foraging behaviour during exposure. Findings are consistent with earlier studies using smaller samples, but the HMM approach provides more nuanced characterization of behaviour changes. 
\end{abstract}
\vspace{0.5em}

\vspace{0.5em}

\section{Introduction}\label{intro}

Measuring and describing behaviour changes by marine mammals in response to underwater noise is a key step in understanding the potential harmful effects of this type of human disturbance on marine ecosystems \citep{Ellison2012,Shannon2015}. The effects of mid-frequency military sonars, airguns used for geophysical exploration, and shipping noise are of particular concern \citep{NationalResearchCouncil2005a, Southall2007}. We focus on potential behaviour changes in response to military mid-frequency active sonar sounds (MFAS) and pseudo-random noise (PRN) in the same frequency range. Here, mid-frequency is broadly 1-10 kHz, but the narrower frequency band containing the primary energy of the MFA and PRN stimuli is between 3-4.5 kHz. 

Controlled exposure experiments (CEEs) are experiments in which animals are tracked -- either visually, or using animal-borne tags -- before, during, and after controlled exposure to the sound stimuli of interest. CEEs are an empirical way to obtain direct measurements of behaviour and potential behavioural responses \citep{Tyack2003a}. Analysis of CEE data then requires qualitative \citep{Miller2012, Sivle2015} or quantitative \citep{Goldbogen2013d, DeRuiter2013a, Antunes2014} analysis of multivariate time series data. These data often include many synoptic variables collected at very high temporal resolution. In some cases, particularly for species of great interest in terms of environmental management, datasets may include only a few individual animals \citep[e.g.,][]{DeRuiter2013,Stimpert2014,Miller2015}.  With rapid technological advances in animal-borne sensors, there have been increases in the amount and complexity of these data. However, the development of quantitative analysis methods for these data have often lagged behind \citep{Holyoak2008}.  In these types of analysis, it is particularly challenging to find methods that combine the relevant data streams in an objective, repeatable way, accounting for individual differences between animals, respecting dependence over time and between streams, and providing output that has a natural interpretation in relation to animal motivation (reasons why an animal performs certain behaviours) and behavioural state. The biological research goal is not simply to provide qualitative descriptions of behaviour, but to conduct hypothesis-driven research into the relationship between acoustic disturbance and behaviour. Given this focus, there is a particular need for models that summarise short-term observations in terms of behavioural states and animal motivation. Such models can facilitate biological interpretation, and also provide inputs for models of Population Consequences of Disturbance (PCoD) \citep{NationalResearchCouncil2005a, Schick2013, New2013a, New2014, King2015}.

Hidden Markov models (HMMs) are well suited to characterising multivariate time series data in terms of a set of latent states \citep{Zucchini2009, Zucchini2008}, which in this case serve as proxies for the underlying behavioural states of the animals.  
HMMs have previously been applied to multivariate animal behaviour data in order to identify latent behavioural states \citep{Bagniewska2013, McKellar2014}, and somewhat-analogous Bayesian hierarchical state-switching models have been used to achieve similar goals \citep{McClintock2013,Isojunno2015}, but no previous work has identified such states in animal-behaviour while accounting for individual differences between tagged animals and incorporating effects of an environmental covariate modulating between-state transition rates.  
Here, we apply HMMs to describe a relatively large dataset on baseline blue whale movement and diving behaviour, including a computationally efficient discrete random effect to account for differences between tag records. The model also incorporates covariates affecting the state transition probability matrix, enabling a quantitative analysis of the potential effects of experimental sounds (MFAS and PRN) on the between-state transition probabilities.

\section{Blue whale CEE response measurements}\label{data}

\subsection{Background: Research program and tags}
Our dataset includes observations of 37 blue whales collected offshore of southern California, U.S.A., as part of the Southern California Behavioral Response Study (SOCAL-BRS). SOCAL-BRS is an interdisciplinary research collaboration designed to study marine mammal behaviour and reactions to sound. Its overall objective is to provide a better scientific basis for estimating risk and minimizing effects of active sonar for the U.S. Navy and regulatory agencies.  The overall experimental methods have been described in detail elsewhere \citep{Southall2012}. However, this study is the first to include MFAS transmissions from operational Navy vessels using full-scale sonar systems within CEEs (detailed below) in addition to simulations of such sonars using transducer arrays deployed from research vessels. 

Whales were tagged with DTAGs \citep{Johnson2003}, animal-borne data loggers that recorded acoustic data (in stereo at 64-240 kHz and 16-bit resolution, with most tags recording at 64 kHz) and high-resolution animal movement data (from tri-axial accelerometers, magnetometers, and a pressure sensor, all sampling at 50-200 Hz). 
Two whales were instead tagged with B-probes (Greeneridge Sciences, Inc., Santa Barbara, CA), which are similar but record acoustic data at up to 20 kHz, with a pressure sensor and 2-axis accelerometers sampling at 1Hz (and no magnetometer).  
The tag data were calibrated, and whale-frame-of-reference depth, acceleration, magnetometer, pitch, roll, and heading data obtained, according to standard methods described in \cite{Johnson2003}. Acoustic records from each tag were processed to obtain flow noise in decibels root-mean-squared (dB RMS re 1 $\mu$Pa) in the 64–94 Hz band;  after application of a 128-order 64-94 Hz bandpass finite impulse response (FIR) filter, the RMS noise level was computed in 1-minute windows sliding forward over the data by 0.2 seconds per measurement, resulting in flow noise data sampled at 5 Hz. Analysed tag recording durations were 1.3-5.6 hours per animal, including 6-93 dives per whale, and 0-16 dives during CEEs per whale (see Table \ref{table-tags} for details of tag deployments). In addition to the tag data, surface observations of animal positions and behaviour were collected by observers following the whales in a small boat.

\begin{table}[!htb]
\caption[Summary of DTAG deployments on blue whales analysed in this study.]{Summary of DTAG deployments on blue whales analysed in this study. Abbreviations used in this table:  ID, identification; h, hours; CEE, controlled exposure experiment; MFAS, mid-frequency active sonar; PRN, psuedo-random noise.} 
\label{table-tags}
\begin{center}
\begin{tabular}{llrrrl}
\hline \hline \\[-2ex]
   \multicolumn{1}{l}{{Whale ID}} &
   \multicolumn{1}{c}{{Date}} &
   \multicolumn{1}{c}{{Dura.\ (h)}} &
   \multicolumn{1}{c}{{Dives}} &
   \multicolumn{1}{c}{{CEE Dives}} &
   \multicolumn{1}{c}{{Exposure}} \\[0.5ex] \hline
   \\[-1.8ex]
  bw10\_235a & 2010-08-23 & 1.4 & 26 &  7 & Simul.\ MFAS \\ 
  bw10\_235b & 2010-08-23 & 1.9 & 31 &  9 & Simul.\ MFAS \\ 
  bw10\_235\_Bprobe\_019 & 2010-08-23 & 1.5 & 24 &  9 & Simul.\ MFAS \\ 
  bw10\_238a & 2010-08-26 & 1.5 & 26 &  9 & Simul.\ MFAS \\ 
  bw10\_239b & 2010-08-27 & 6.0 & 53 &  5 & Simul.\ MFAS \\ 
  bw10\_240a & 2010-08-28 & 1.6 & 54 & 16 & Simul.\ MFAS \\ 
  bw10\_240b & 2010-08-28 & 2.7 & 16 &  3 & Simul.\ MFAS \\ 
  bw10\_241\_Bprobe\_034 & 2010-08-29 & 2.3 & 27 &  8 & Silent \\ 
  bw10\_241a & 2010-08-29 & 2.9 & 16 &  3 & Silent \\ 
  bw10\_243a & 2010-08-31 & 4.8 & 22 &  3 & PRN \\ 
  bw10\_243b & 2010-08-31 & 4.4 & 22 &  3 & PRN \\ 
  bw10\_244b & 2010-09-01 & 1.6 & 11 &  2 & PRN \\ 
  bw10\_244c & 2010-09-01 & 2.4 & 17 &  6 & PRN \\ 
  bw10\_245a & 2010-09-02 & 5.2 & 36 &  3 & PRN \\ 
  bw10\_246a & 2010-09-03 & 4.0 & 27 &  2 & Simul.\ MFAS \\ 
  bw10\_246b & 2010-09-03 & 4.4 & 15 &  2 & Simul.\ MFAS \\ 
  bw10\_251a & 2010-09-08 & 3.0 & 26 &  7 & PRN \\ 
  bw10\_265a & 2010-09-22 & 5.2 & 44 &  3 & Simul.\ MFAS \\ 
  bw10\_266a & 2010-09-23 & 2.5 & 17 &  5 & PRN \\ 
  bw11\_210a & 2011-07-29 & 3.3 & 17 &  3 & Simul.\ MFAS \\ 
  bw11\_210b & 2011-07-29 & 4.9 & 33 &  5 & Simul.\ MFAS \\ 
  bw11\_211a & 2011-07-30 & 3.7 & 24 &  2 & PRN \\ 
  bw11\_213b & 2011-08-01 & 4.9 & 51 &  6 & Simul.\ MFAS \\ 
  bw11\_214b & 2011-08-02 & 4.9 & 93 & 10 & PRN \\ 
  bw11\_218a & 2011-08-06 & 2.6 & 17 &  4 & PRN \\ 
  bw11\_218b & 2011-08-06 & 3.6 & 23 &  2 & PRN \\ 
  bw11\_219b & 2011-08-07 & 2.1 & 10 &  0 & None \\ 
  bw11\_220a & 2011-08-08 & 1.1 &  6 &  3 & Simul.\ MFAS \\ 
  bw11\_220b & 2011-08-08 & 2.1 & 15 &  6 & Simul.\ MFAS \\ 
  bw11\_221a & 2011-08-09 & 4.3 & 33 &  4 & PRN \\ 
  bw11\_221b & 2011-08-09 & 2.8 & 17 &  4 & PRN \\ 
  bw12\_292a & 2012-10-18 & 4.4 & 14 &  3 & PRN \\ 
  bw13\_191a & 2013-07-10 & 4.8 & 41 &  6 & Real MFAS \\ 
  bw13\_207a & 2013-07-26 & 3.4 & 37 &  8 & Silent \\ 
  bw13\_214b & 2013-08-02 & 3.5 & 21 &  0 & None \\ 
  bw13\_217a & 2013-08-05 & 2.3 & 13 &  0 & None \\ 
  bw13\_259a & 2013-09-16 & 5.4 & 67 &  7 & Simul.\ MFAS \\ 
   \hline
\end{tabular}
\end{center}
\end{table}

\subsection{Controlled exposure experiment protocols}
Data were collected before, during, and after CEEs, in which whales were exposed to either MFAS or PRN sounds. The MFAS signals were either real (from a U.S. Navy vessel) or simulated (from a custom-built underwater transducer array). MFAS sounds included several tonal and frequency upsweep signals with frequencies between 3.2-3.6 kHz (real) or 3.5-4.05 kHz (simulated). The second signal type was PRN, which was band-limited noise in the 3.5-4.05 kHz band. PRN signal duration and frequency content was similar to MFAS, but lacking the tonal and FM upsweep component. Simulated MFAS and PRN CEEs each lasted approximately 30 minutes, but the single real MFAS experiment was 58 minutes. During the real MFAS CEE, the distance to the subject was much greater in order to match the predicted received level for the smaller sources, and a longer duration was required to cover a comparable experimental range. 

Each CEE consisted of multiple sequential signals presented at the nominal repetition rate of the common MFAS systems whose potential effects motivated the study. During all simulated MFAS and PRN CEEs, individual experimental stimuli (about 1.5 seconds in duration) were transmitted approximately once every 25 seconds, with an initial source level of 160 dB re 1 µPa RMS at 1 m. The source level was increased by 6 dB per transmission until the maximum source level (210 dB re 1 $\mu$Pa RMS at 1 m for MFA or 206 for PRN) was reached; transmissions then continued at maximum level until the conclusion of the CEE. For the single real MFAS exposure, such a ramp-up was not feasible and is not part of normal operations. All transmissions were at the nominal source level for the SQS-53C sonar operated on the Navy ship participating in the study: 235 dB re 1 μPa RMS at 1m. A repetition rate of one signal every 25s was used. The distance from the real ship to the subject whale in this case was prescribed as an approach from about 22-12 km, whereas the experimental source transmitting simulated MFAS and PRN signals was both much closer (nominally 1-2 km) and was stationary during CEEs.

After CEEs, tag data recording continued until the tag detached from the animal (usually at least one hour, except in case of unplanned early detachment), and surface observations were also maintained for at least an hour post-CEE (except when prevented by darkness or when the animal was lost). While the main vessel (which operated the sound source) had somewhat variable orientation with respect to the animal, a smaller boat typically remained within several hundred meters of the whale during the entire observation period in order to collect the visual observations (such as precise locations of the whale's surface positions), but also to maintain a consistent context for behavioural observations \citep{Southall2012}. 

\subsection{Whale behaviour data}
Using the high-resolution, multivariate tag data and the the visual observations of behaviour and spatial locations, a number of variables were chosen to summarise the whales' behaviour.  These variables were computed on a dive-by-dive basis; in other words, the input data for modelling were time series for which the sampling unit was one dive.  Here, a ``dive" was defined as any excursion from the surface to 10 m depth or greater. Dive start- and end-times were detected in the dive profile by visual inspection. The variables calculated for each dive were dive duration, post-dive surface duration, maximum depth, step length and turning angle in the horizontal dimension, number of feeding lunges, and variability of heading. Dive duration was the time (in seconds) from the start of a dive until the first surfacing following the dive.  The post-dive surface duration was defined as the time (in seconds) from the end of one dive until the start of the subsequent dive.  Maximum dive depth was the maximum depth attained by the whale over the course of the dive.  To compute step lengths and turning angles based on position observations collected by human observers, we first used linear interpolation on the visual observation data on whale positions, which were unevenly spaced in time.  The resulting interpolated tracks had a position estimate at the mid-point time of each dive.  We verified the suitability of this simple track interpolation method by visual comparison of the original and interpolated tracks for all whales (data not shown).  

\begin{figure}
\centering
\includegraphics[width=\textwidth]{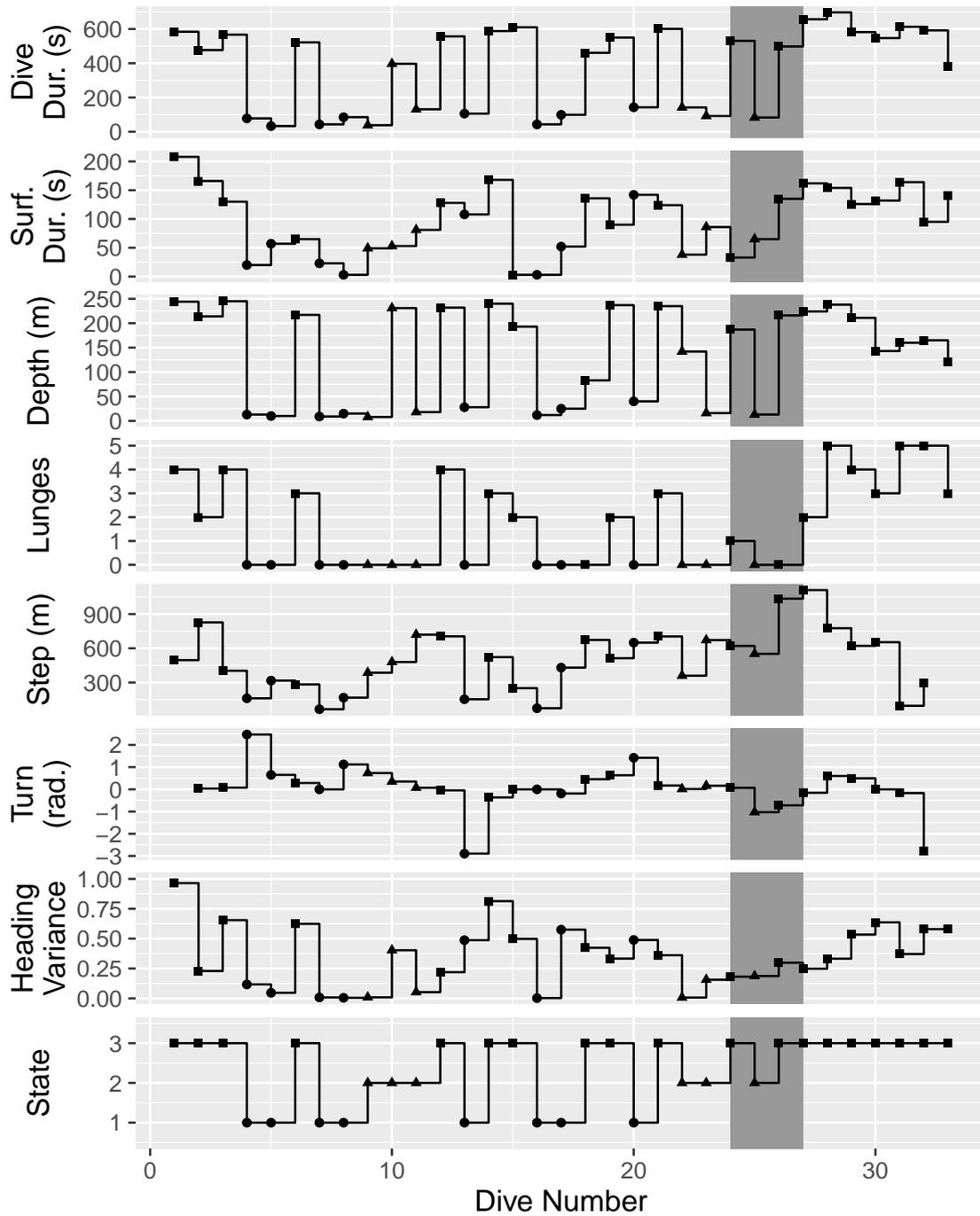} \caption{\label{fig-data1} Time-series plot of the input data for whale bw11\_221a. The CEE exposure period is shaded in darker grey; this whale was exposed to
a pseudo-random noise (PRN) CEE. The bottom panel shows the most probable state for each dive, according to the best model (with 3 states, 4 contexts, and a common effect of acoustic disturbance). These states are also indicated in the other panels by symbols: circles for state 1, triangles for state 2, and squares for state 3.}
\end{figure}

\begin{figure}
\centering
\includegraphics[width=\textwidth]{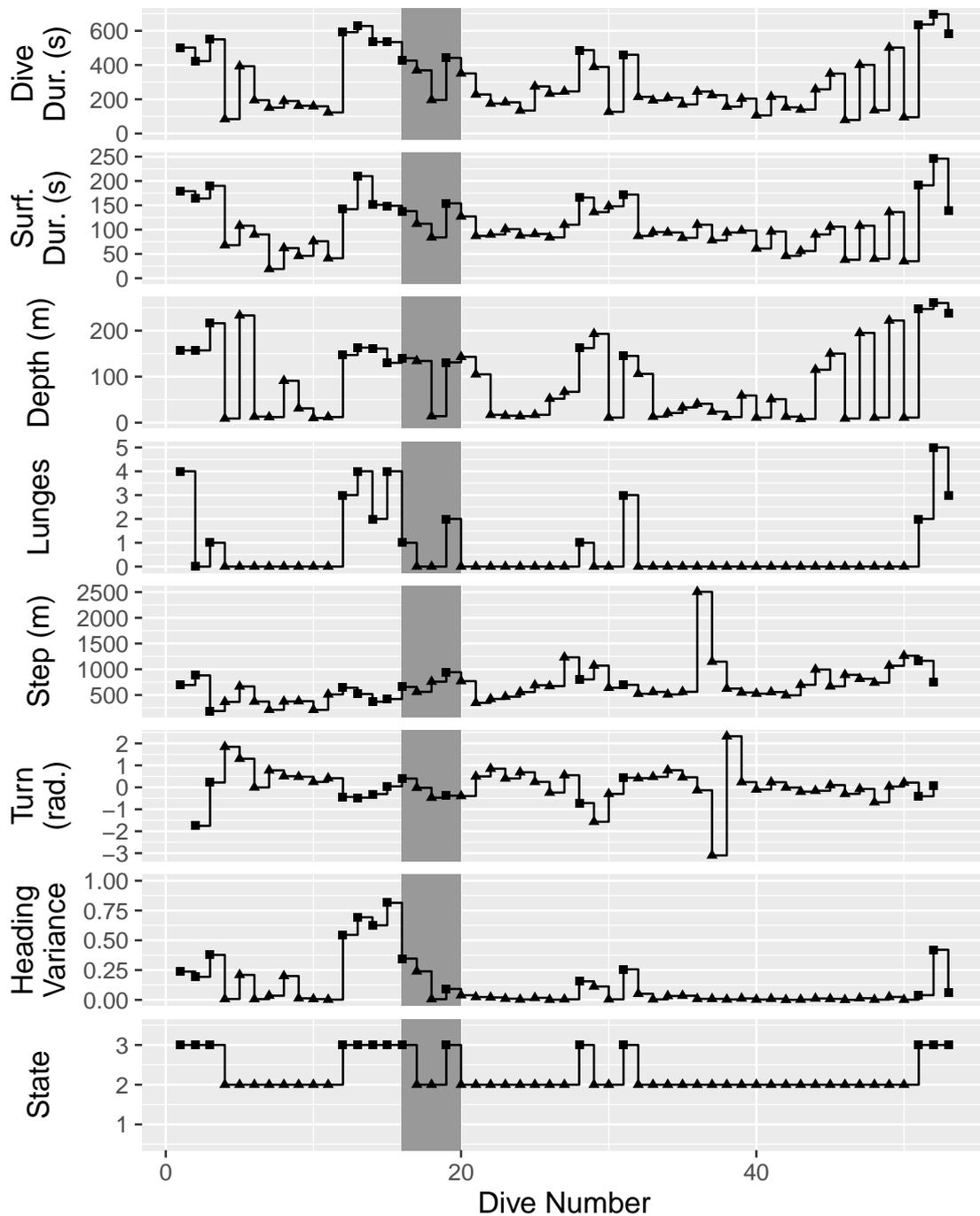} \caption{\label{fig-data2} Time-series plot of the input data for whale bw10\_239b. The CEE exposure period is shaded in darker grey; this whale was exposed to a simulated MFA sonar CEE. The bottom panel shows the most probable state for each dive, according to the best model (with 3 states, 4 contexts, and a common effect of acoustic disturbance). These states are also indicated in the other panels by symbols: circles for state 1, triangles for state 2, and squares for state 3.}
\end{figure}

Finally, we computed step length and turning angle based on the interpolated tracks. Step length was the rhumb-line distance travelled (in metres) between the mid-point of a given dive and that of the subsequent dive.  The turning angle was the heading angle change (in radians) comparing the previous animal heading (defined by the vector pointing from the previous position to the current position) and the current animal heading (defined by the vector pointing from the current position to the subsequent position).  The number of feeding lunges during each dive was determined by visual inspection of the tag depth, accelerometer, and animal orientation data, along with low-frequency flow noise levels from the tag acoustic recording; lunges were characterised by the distinctive combination of a vertical excursion in the dive profile (generally during the bottom phase of the dive) together with characteristic body pitch and a sudden deceleration evident in the body acceleration and flow noise data (here, flow noise intensity from the tag acoustic record served as a proxy for swim speed) \citep{Goldbogen2006}.  Finally, variability of heading was computed from the DTAG heading data, using the circular variance metric, as appropriate for angular data \citep{Zar2010}; this variance ranges from 0 (no variation) to 1 (random variation in angles between 0 and 2$\pi$).
Figures \ref{fig-data1}-\ref{fig-data2} show example time series plots of the data for two of the 37 whales. 

\section{Modelling the state-switching behaviour using HMMs}\label{}

\subsection{A baseline model}

\subsubsection{Motivation}

The data suggest that at least two (and probably three) distinct behavioural states drive the magnitude of the observations made \citep{Goldbogen2013}. For example, the deepest, longest dives usually contain multiple lunges, and represent deep feeding behaviour; there are also shorter, shallow feeding dives that usually contain a single lunge (feeding event); finally, some relatively short, shallow dives without lunges have very low heading variance, and may represent near-surface traveling. Furthermore, there is autocorrelation in the observed time series, clearly due to persistence of behavioural states, with animals tending to repeat the same type of dive several times before switching to a different behavioural state. 

HMMs naturally accommodate both these features. A basic $N$-state HMM in discrete time involves two components, 1) an observed state-dependent process and 2) an unobserved $N$-state Markov chain, with the observations of the state-dependent process assumed to be generated by one of $N$ component distributions corresponding to the $N$ states of the Markov chain. In HMM applications to animal behaviour data, the states of the Markov chain can often naturally be interpreted as proxies for the behavioural states of animals, although there is not necessarily a one-to-one correspondence between nominal HMM states and biologically meaningful behavioural states \citep{Zucchini2008,Langrock2012,Langrock2015}.

\subsubsection{Formulation of the initial model}\label{initial-model}

In the given application, for each whale, the observed state-dependent process is multivariate and will be denoted by $\{ \mathbf{X}_{wd\cdot} \}_{d=1,2,\ldots, D_w}$, where $\mathbf{X}_{wd\cdot} = (X_{wd1}, \ldots, X_{wdP})$ is the vector of variables observed for dive $d$ performed by whale $w$, and $D_w$ is the total number of dives performed by that whale. In our application, we have $P=7$ data streams. The $N$-state Markov chain giving the (behavioural) states that underlie the dives performed by whale $w$ will be denoted by $\{ S_{wd} \}_{d=1,2,\ldots, D_w}$, where ${S}_{wd}$ gives the state associated with dive $d$. Thus, with our model we classify dives into a finite number of categories \citep{Hart2010,Bagniewska2013}, rather than modelling diving activity at a finer, within-dive scale \citep{Langrock2014}. 

We assume a basic first-order dependence structure for the state process, and, for whale $w$, summarize the state transition probabilities $\gamma_{ij}^{(w)}=\Pr (S_{wd} = j \mid S_{w,d-1} = i)$ in the transition probability matrix (t.p.m.)
$$ \boldsymbol{\Gamma}^{(w)}= (\gamma_{ij}^{(w)}) = \begin{pmatrix} \gamma_{11}^{(w)} & \ldots & \gamma_{1N}^{(w)} \\ \vdots & \ddots & \vdots \\  \gamma_{N1}^{(w)} & \ldots & \gamma_{NN}^{(w)} \end{pmatrix}.$$
We initially assume that $\boldsymbol{\Gamma}^{(w)}=\boldsymbol{\Gamma}=(\gamma_{ij})$ for all $w$ (i.e., complete pooling: all whales have the same state-switching dynamics), but this assumption will be relaxed later on. For whale $w$, the initial state distribution is denoted by $\boldsymbol{\delta}^{(w)}=\bigl( \Pr (S_{w1} = 1), \ldots , \Pr (S_{w1} = N) \bigr) $, again initially assuming that $\boldsymbol{\delta}^{(w)}=\boldsymbol{\delta}$ for all $w$.

We plan to use an unconstrained numerical optimizer to find the maximum likelihood estimates, and hence reparameterize all model components in terms of unconstrained parameters. In particular, we let
$$ \gamma_{ij} = \frac{\exp (\alpha_{ij}) }{ 1 + \sum_{l \neq i } \exp (\alpha_{il}) }, $$
and perform the maximization of the likelihood with respect to the real-valued parameters $\alpha_{ij}$, $i,j=1,\ldots,N$, $i \neq j$. This reparameterization based on the multinomial logit function will be convenient also when extending the model to incorporate random effects and covariates (see subsequent sections).

We assume that the different time series observed, corresponding to the different individual whales, are independent of each other. For the state-dependent processes, we assume that given the state underlying dive $d$, the observation vector $\mathbf{X}_{wd\cdot}$ is conditionally independent of past and future observations and states. Furthermore, for the observation vector $\mathbf{X}_{wd\cdot}$ we assume {\em contemporaneous conditional independence}, given the states, i.e.,
$$ f(\mathbf{x}_{wd\cdot} \mid {S}_{wd} = {s}_{wd}) = \prod_{p=1}^P f({x}_{wdp} \mid {S}_{wd} = {s}_{wd}); $$
Note that we use $f$ as a general symbol for a density or probability function. For more details on the assumption of contemporaneous conditional independence, see \citet{Zucchini2009}.  
Note that, unconditionally, the $P=7$ component variables will still be correlated with each other, because the Markov chain induces dependence between them. 

As a consequence of the contemporaneous conditional independence assumption, to formulate a model for the blue whale dataset, we simply need to specify one univariate distribution for each of the seven variables observed. (The parameters of each distribution then depend on the underlying state.) We model the dive depth, dive duration, post-dive surface duration and step length with gamma distributions, since those variables all take on positive real values and may have right-skewed distributions.  For the gamma distributions, we employ a parameterisation in terms of the mean $\mu$ and standard deviation $\sigma$; if required, these parameters can be easily converted to the more standard shape ($k$) and scale ($\theta$) parameterisation according to $k = \mu^2/\sigma^2$ and $\theta = \sigma^2/\mu$. For the number of lunges, we use a Poisson distribution, a standard choice for count data.  We select a von Mises distribution --- which is a circular analogue of the normal distribution --- for the turning angle data.  Finally, we model the heading variance with a beta distribution, which, like the data, has support on the interval $[0,1]$.

\subsubsection{Model fitting}

A key property of HMMs is that an efficient dynamic programming algorithm, the forward algorithm, can be used to evaluate the likelihood. The forward algorithm exploits the dependence structure of the HMM to calculate the likelihood recursively, traversing along the time series and updating the likelihood at every step, while maintaining the current state probabilities. This renders it computationally feasible and convenient to estimate the model parameters by numerically maximizing the likelihood \citep{MacDonald2014}. For the model specified above --- with both t.p.m., $\boldsymbol{\Gamma}$, and initial state distribution, $\boldsymbol{\delta}$, being identical across individuals --- the use of the forward algorithm leads to the likelihood expression
\begin{equation}\label{like1}
\mathcal{L} = \prod_{w=1}^{37} \biggl( \boldsymbol{\delta} \mathbf{Q}(\mathbf{x}_{w1\cdot}) \boldsymbol{\Gamma} \mathbf{Q}(\mathbf{x}_{w2\cdot}) \cdot \ldots \cdot \boldsymbol{\Gamma} \mathbf{Q}(\mathbf{x}_{wD_w\cdot}) \mathbf{1} \biggr),
\end{equation}
where $D_w$ denotes the number of dives whale $w$ performed during the observation period, $\mathbf{Q}(\mathbf{x}_{wd\cdot})$ is an $N \times N$ diagonal matrix with the $k$-th entry on the diagonal given by $\prod_{p=1}^P f({x}_{wdp} \mid {S}_{wd} = k)$ (i.e., the joint density of the observations ${x}_{wd1}, \ldots,  {x}_{wdp}$, given state $k$), and $\mathbf{1} \in \mathbb{R}^N$ is a column vector of ones. 

We note that an advantage of the HMM formulation described above is its ability to deal with missing and partially missing data. In ecological datasets in particular, on some occasions data may be available from some variables in the observed multivariate state-dependent process, but missing for other variables.  For example, in the current dataset, observations for some whales span nightfall.  After dark, surface visual observations ceased, so all data derived from visually-observed positions are missing.  However, the tag-derived variables are recorded as usual regardless of darkness.  As another example, some whales in the study were tagged with B-probes rather than DTAGs.  The B-probe devices do not have tri-axial accelerometers and magnetometers, so they do not collect any heading data.  An HMM --- with an assumption of contemporaneous conditional independence, as detailed earlier in Section \ref{initial-model} --- can easily accommodate such partially observed data in the state-dependent process.  If any $x_{wdp}$ (the observation of variable $p$ for dive $d$ by whale $w$) is missing, we simply have $f(x_{wdp} \mid S_{wd}=k)=1$ in Equation (\ref{like1}) \citep{Zucchini2009}; in consequence, the missing data point makes no contribution to $\mathbf{Q}(\mathbf{x}_{wd\cdot})$ and thus does not affect the overall likelihood, but observations of other variables for that same dive do still contribute.

For the blue whale dataset, in the case of three states ($N=3$), it took five minutes to numerically maximize this model's likelihood using \texttt{nlm()} in R \citep{RCoreTeam2015} (on an octa-core i7 CPU, at 2.7 GHz and with 4 GB RAM).

\subsubsection{Results and interpretation of the states}

\setlength{\tabcolsep}{4pt}
\begin{footnotesize}
\begin{table}[!htb]
\caption{Parameter estimates for the state-dependent distributions for the 3-state model with complete pooling.  For each variable, the distribution fitted to the data is indicated in parentheses.}
\label{table-3state-est}
\begin{center}
\begin{tabular}{lcccc}
       Variable              & State 1  & State 2 &  State 3 \\ \hline
Duration  (gamma)            & $\mu$ = 135.4, $\sigma$ = 75.3 & $\mu$ = 350.5, $\sigma$ = 216.1 & $\mu$ = 508.2, $\sigma$ = 135.8 \\
Surf.\ Duration (gamma)     & $\mu$ = 70.5, $\sigma$ = 68.4  & $\mu$ = 86.4, $\sigma$ = 53.9 & $\mu$ = 148.3, $\sigma$ = 69.1 \\
Max.\ Depth (gamma)        & $\mu$ = 30.5, $\sigma$ = 21.8  & $\mu$ = 71.3, $\sigma$ = 67.2 & $\mu$ = 166.7, $\sigma$ = 62.0 \\
No.\ Lunges (Poisson)   & $\lambda$ = 0.60  & $\lambda$ = 0.01 & $\lambda$ = 3.30 \\
Step Length (gamma)          & $\mu$ = 193.8, $\sigma$ = 139.1  & $\mu$ = 710.7, $\sigma$ = 294.4 & $\mu$ =401.3, $\sigma$ = 281.2 \\
Turn.\ Angle (von Mises)    & $\mu$ == 0, $\kappa$ = 1.04 & $\mu$ == 0, $\kappa$ = 3.11 & $\mu$ == 0, $\kappa$ = 0.83\\
Head.\ Variance (beta)      & $a$ = 0.88, $b$ = 2.05 & $a$ = 0.52, $b$ = 6.16 & $a$ = 1.68, $b$ = 1.59 \\
\hline
\end{tabular}
\end{center}
\end{table}
\end{footnotesize}

\begin{figure}[!htb]
\centering
\makebox{\includegraphics[width=\textwidth]{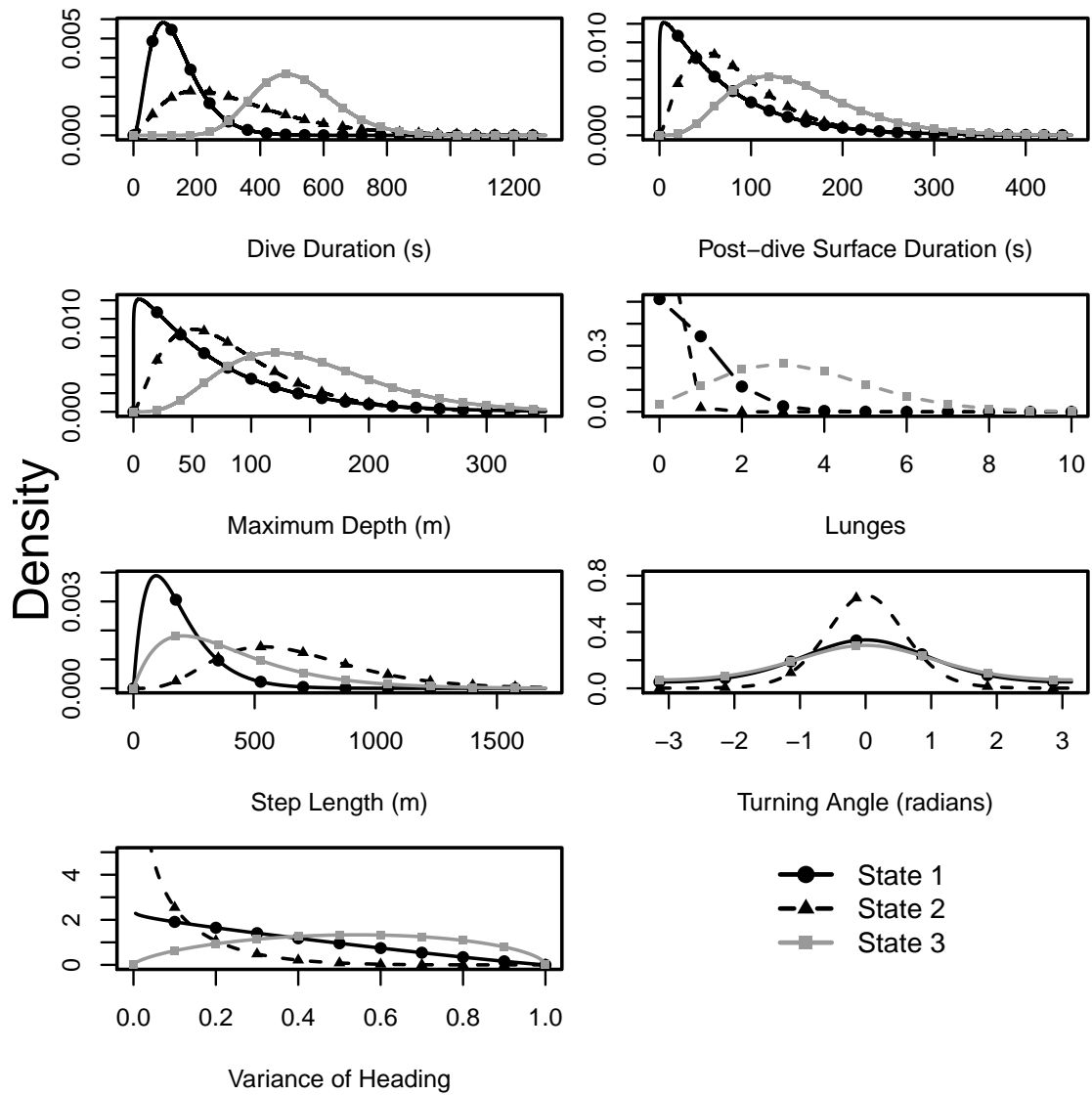}} \caption{\label{fig-state-dists} Fitted model estimates of the state-dependent distributions for the 3-state HMM with complete pooling.}
\end{figure}

The resulting parameter estimates are presented in Table \ref{table-3state-est} and Figure \ref{fig-state-dists}. The t.p.m.\ was estimated as
$$ \hat{\boldsymbol{\Gamma}} =
\begin{pmatrix}
 0.931 & 0.014 & 0.055 \\
 0.018 & 0.785 & 0.197 \\
 0.071 & 0.100 & 0.829
\end{pmatrix}
$$
The corresponding stationary distribution is ${\boldsymbol{\hat{\delta}}}=(0.433,0.199,0.368)$.

This simple model succeeds in capturing several of the important features of the data. In particular, it identifies three biologically relevant dive types.  State 1 includes shorter, shallower dives with 0-1 lunges per dive, short step length and variable turning angle and heading variance; these are likely shallow foraging dives.  State 2 comprises dives of moderate length and depth, without lunges, and with long step lengths and small turning angles and heading variance; these probably represent directed travel by the whales (without feeding).  Finally, state 3 likely includes deep foraging dives, with long duration, deep depth, many lunges, moderate step length, and variable turning angle and heading variance.  These three dive types provide an adequate, if simplistic, classification of the main dive types performed by all the whales in the data set. 

In all models from this point forward, we consider only formulations with three states.  Although this is a somewhat arbitrary choice, the selected model is both biologically informative and interpretable, serving the desired scientific purpose, and is generally consistent with previous classifications of blue whale behavioural states \citep{Goldbogen2013c}.  In this case, it is likely that formal model selection for the number of states would lead to a very high number of states, resulting in a model that would be essentially uninterpretable \citep{Langrock2015}. Therefore, we chose to focus on three-state models for the practical reasons mentioned.

Although the initial three-state model captures several important features of the data, in the raw data, it is immediately evident that the proportion of dives of each type, and the transition rates between them, vary strongly from whale to whale (see Figures \ref{fig-data1}-\ref{fig-data2} and web-based supporting materials).  Accordingly, as the next step in model development, we chose to relax the assumption of homogeneity between whales (tag records).

\subsection{Accounting for heterogeneity}

\subsubsection{Motivation}

Comparing the time-series dive data between whales, several clear patterns indicate that the assumption of homogeneity among individuals is inadequate. Some whales have one predominant dive type. For example, whales bw10\_243a, bw10\_243b, bw10\_244b and bw13\_217a perform almost solely deep, multi-lunge dives, while bw10\_235a, bw10\_239b and bw10\_251a perform numerous consecutive shallow dives without lunges and with larger turning angles. Other whales alternate more regularly between dive types.  For example, bw10\_235b, bw10\_240a, bw10\_266a, bw11\_214b and bw11\_221a alternate dives with and without lunges, without very long series of either.  

This variation may be evidence of genuine individual differences between whales, although that conclusion is difficult to confirm without re-tagging the same individual multiple times in different situations \citep{Goldbogen2013d}.  Even in the absence of individual differences as such, it is of biological interest to account for differences between tag recordings.  Blue whale behaviour, particularly feeding behaviour, is strongly driven by environmental covariates such as the density and depth at which prey are present \citep{Hazen2015, Goldbogen2013}.  These covariates are very difficult to measure, and are missing from many observations of whale behaviour (including those of most individuals in our study). Variation in prey abundance and depth, among other environmental covariates, may drive whales to adopt a corresponding behavioural ``context" \citep{Friedlaender2015a}.  In other words, blue whale forage differently depending on prey (krill) abundance: when prey are shallow and more dispersed, whales feed with low lunge counts and maximize oxygen conservation; when deep, dense krill patches are available, whales likely go into oxygen debt in order to lunge as many times as possible and maximize energy gain \citep{Goldbogen2015, Hazen2015}.  The increased maneuvering necessary to harvest low-prey-density patches relative to dense ones is also significant \citep{Goldbogen2015, Hazen2015}. Within this framework, it seems reasonable to assume that there is a finite, relatively small number of behavioural contexts adopted by blue whales in our study region.    

\subsubsection{Formulation of a model with discrete-valued random effects}

\citet{Maruotti2009} and \citet{McKellar2014} suggest the use of discrete-valued random effects in HMMs to capture this type of heterogeneity across component time series. This approach has two main advantages over the more common use of continuous-valued random effects: (1) the computational effort is substantially reduced, and (2) restrictive distributional assumptions on the random effects are avoided. Furthermore, interpretation is often more intuitive than for continuous-valued random effects. For example, each outcome of a discrete random effect incorporated in the t.p.m.\ corresponds to one type of state-switching pattern, which in turn corresponds to one of the behavioural contexts described earlier.

To incorporate discrete random effects into our blue whale dive behaviour model, we assume that
$$ \boldsymbol{\Gamma}^{(w)} = \boldsymbol{\Gamma}_{\xi^{(w)}} = \bigl(\gamma_{ij\xi^{(w)}}\bigr) $$
with an i.i.d.\ discrete random effect $\xi^{(w)}$ such that $\Pr(\xi^{(w)}=k)=\pi_k$ for $k=1,\ldots,K$, and
$$ \gamma_{ijk} = \Pr \bigl( S_{wd} = j \mid S_{w,d-1} = i, \xi^{(w)}=k  \bigr) = \frac{\exp (\alpha_{ijk}) }{ 1 + \sum_{l \neq i } \exp (\alpha_{ilk}) }, $$
for $k=1,\ldots,K$. With this mixture model for the state-switching dynamics, there are $K$ possible $\boldsymbol{\Gamma}$s, $\boldsymbol{\Gamma}_1,\ldots,\boldsymbol{\Gamma}_K$, and each individual whale's time series is assumed to be driven by exactly one of them. The mixture weight $\pi_k$, $k=1,\ldots,K$, gives the probability that the t.p.m.\ $\boldsymbol{\Gamma}_{k}$ underlies the state-switching dynamics for a given whale. The number of mixture components, $K$, controls the flexibility of the model to account for heterogeneity and can be chosen using model selection criteria. The initial state distributions are also assumed to be context-specific, i.e., $ \boldsymbol{\delta}^{(w)} = \boldsymbol{\delta}_{\xi^{(w)}}$, where the $K$ different initial state distributions, $\boldsymbol{\delta}_{1}, \ldots, \boldsymbol{\delta}_{K}$, are estimated alongside the other model parameters.

\subsubsection{Model fitting}\label{fancyfit}

The likelihood of the above HMM with discrete-valued random effects is 
\begin{equation}\label{like2}
\mathcal{L} = \prod_{w=1}^{37} \sum_{k=1}^K \mathcal{L}_{k}^{(w)} \pi_k = \prod_{w=1}^{37} \sum_{k=1}^K \biggl( \boldsymbol{\delta}_k \mathbf{Q}(\mathbf{x}_{w1\cdot}) \boldsymbol{\Gamma}_k \mathbf{Q}(\mathbf{x}_{w2\cdot}) \cdot \ldots \cdot \boldsymbol{\Gamma}_k \mathbf{Q}(\mathbf{x}_{wD_w\cdot}) \mathbf{1} \pi_k \biggr).
\end{equation}
Here $\mathcal{L}_{k}^{(w)}$ denotes the conditional likelihood of the observations made for whale $w$ given the $k$-th behavioural context (i.e.\ given $\xi^{(w)}=k$ and hence $\boldsymbol{\Gamma}^{(w)} = \boldsymbol{\Gamma}_{k}$); otherwise the notation is the same as in (\ref{like1}).

Fitting this more complicated model via naive application of a numerical optimizer presented several difficulties. Due to the fairly large number of parameters, the optimisation was slow and subject to numerical instability in terms of identification of local rather than global maxima of the likelihood. To allow for an efficient fitting of the model, we used several complementary strategies. First, we wrote the HMM forward algorithm using the Rcpp package \citep{Eddelbuettel2013}, which provides a simple way to use C++ code in conjunction with R software. This substitution of C++ code for R code greatly reduced the computation time required for model fitting. Second, we carefully selected appropriate starting values following a two-stage procedure: 

I. In the first stage, we determined reasonable starting values for the parameters of the state-dependent distributions. Initial exploratory analysis indicated that these parameter values are stable for any number of states $N$ and any number of possible outcomes of the discrete random effect $K$. Thus, we ran the optimizer for $N=3$ and $K=1$ and used the resulting estimates of state-dependent distribution parameters as starting values for the rest of the optimization stages. 

II. In the second stage of model fitting, the goal was to find reasonable initial values for the parameters determining the Markov state process, i.e.\ the entries of the transition probability matrices $\boldsymbol{\Gamma}_k$ and the corresponding initial distributions $\boldsymbol{\delta}_k$. Initial exploration indicated that the estimators of these parameters are highly unstable, so selecting good starting values is crucial. We fixed the state-dependent distribution parameters at their initial values, and allowed only the parameters of the state process to vary during this optimisation stage. (Fixing the parameters of the state-dependent distributions decreased the required computing time by about a factor of 20.) To select initial values for the parameters of the state process, we ran the optimization 15000 times, using random starting values in each run. The resulting sets of Markovian parameter estimates were ranked in terms of their corresponding data likelihood, and the best 100 sets were used as initial values for the next stage of model fitting (comprehensive optimisation). 

Finally, we carried out a full numerical maximisation of the likelihood. Initially, we simply ran the numerical maximiser, 
selecting the initial values for the state-dependent distribution parameters and the Markov chain parameters as described earlier. 
To further decrease the chance of finding only local maxima rather than the global maximum, we then additionally jittered the parameter estimates and re-ran the optimiser. Specifically, for each of the 100
best models identified in the previous stage, we added a small amount of noise 
to each parameter estimate, and used these jittered values as the initial values to re-run the optimiser. This was done five times for each of the 100 best models identified. This jittering procedure is similar to a bootstrap restart as suggested by \citet{Wood2001}; we also implemented the latter, but for our data it did not yield any improvements.

\subsubsection{Model selection on the number of behavioural contexts}
All of the sets of parameter estimates obtained were ranked according to their corresponding likelihoods, and we selected the maximum likelihood estimates to define the best model for each value of $K$. We considered $K$ (the number of behavioural contexts) to take on candidate values from 1 to 5, selecting the optimal $K$ based on AIC scores.

\begin{table}
\caption{Log-likelihood and AIC values for the different models considered.  The last column also gives the difference in AIC for each model relative to the base model (3 states, one behavioural context, and no effect of acoustic exposure).}
\label{table-AICs} 
\begin{center}
\begin{tabular}{cccccr}
\hline\\[-0.7em]
type of                       & $K$  & number of  & $\log \mathcal{L}$ & AIC            & $\Delta$ AIC   \\

model                         &      & parameters &                    &                &                \\[0.4em]
\hline\\[-0.7em]
no effect of 
acoustic exposure             & 1    & 44         & -25736.6           &  51561.1       &      0         \\
                              & 2    & 53         & -25682.9           &  51471.8       &    -89.3       \\
                              & 3    & 62         & -25665.4           &  51454.8       &   -106.3       \\
                              & 4    & 71         & -25651.0           &  51444.0       &   -117.1       \\
                              & 5    & 80         & -25643.3           &  51446.5       &   -114.6       \\[0.4em]
\hline\\[-0.7em]
common acoustic 
exposure effect               & 4    & 77         & -25642.5           &  {\bf 51438.9} &   -122.2      \\[0.4em] 
\hline\\[-0.7em]
context-specific acoustic 
exposure effect               & 4    & 95         & -25636.8           &  51463.7       &    -97.4      \\[0.4em] 
\hline
\end{tabular}
\end{center}
\end{table}

Table \ref{table-AICs} provides model selection results, including log-likelihoods and AIC values for the various models considered.  According to the AIC scores, the optimal number of behavioural contexts, $K$, is four for these blue whale data.  
Rather than reporting model parameter estimates at this point, we defer those results to the next section, where we consider one final refinement of the model: inclusion of an effect of acoustic exposure on the transition probability matrices.  

\subsection{Investigating the effect of sound exposure}

\subsubsection{Motivation}

As detailed in Section \ref{data}, most of the whales included in our dataset were subject to CEEs in which they were exposed to PRN or MFAS sounds (real or simulated).  Previous analysis of a subset of the current dataset has suggested that blue whale behavioural responses to CEEs did not uniformly occur but that when they did they were variable, complex, and included responses such as the cessation of deep feeding behaviour and initiation of directed travel \citep{Goldbogen2013d}.  To allow for such responses, we modified our model formulation to include the effect of an acoustic exposure covariate on the transition probability matrix (or matrices). We note that models such as this one, which incorporate both random effects and covariates, fall into the category of mixed HMMs as discussed in \citet{Altman2007}. 

Although the number of CEEs included in our dataset is quite large compared to all other such studies, it is still small from a statistical perspective, so we did not attempt to differentiate between exposures of different types (14 simulated MFA, 1 real MFA, and 14 PRN CEEs).  Rather, we sought to assess the utility of HMMs to investigate potential changes in blue whale state-switching behaviour in response to these somewhat similar mid-frequency sounds. Previous studies \citep{Goldbogen2013c} and ongoing analyses are considering individual responses by specific sound type to compare potential differential responses, including identifying changes as a function of received sound level. Here, we assumed that the effects of all CEE sounds were identical regardless of sound type and received sound exposure level, and that silent control CEEs had no effect on behaviour.  While these assumptions undoubtedly represent simplifications of reality, the results of previous work suggest that they should probably be approximately appropriate for this species \citep{Goldbogen2013d}.

\subsubsection{Model formulation of a model that incorporates sound exposure}

In models with an acoustic exposure covariate, altered t.p.m.s are in place during times when whales were subject to sonar exposure.  For each behavioural context, the between-state transition probabilities $\gamma_{ijkd}^{(w)}=\Pr \bigl( S_{wd} = j \mid S_{w,d-1} = i, \xi^{(w)}=k  \bigr)$ are now defined as
$$ \gamma_{ijkd}^{(w)} = \frac{\exp \bigl( \alpha_{ijk} + \beta_{ijk}   z_{wd} \bigr) }{ 1 + \sum_{l \neq i } \exp \bigl(\alpha_{ilk}  + \beta_{ilk} z_{wd} \bigr) }, $$
where
$$ z_{wd} =
\begin{cases}
1 & \text{if whale $w$ was exposed to sonar during dive $d$;} \\
0 & \text{otherwise.} \\
\end{cases} $$
In other words, for each of the $K$ contexts, there are two t.p.m.s, one associated with baseline (undisturbed) periods, and the other with CEE (potentially disturbed) periods.

We consider several possible constraints on the $\beta$ parameters, which quantify the strength and direction of the relationships between exposure and each transition probability.  In the most general case, the effect of acoustic exposure is context-specific, so that each $\beta_{ijk}$ has a unique value for each $k$, $k=1,\ldots,K$; the effect of exposure is different for each t.p.m. entry within every behavioural context.  We also consider the possibility of a common acoustic exposure effect, where the $\beta_{ijk}$ have constant values across all contexts, i.e.\ $\beta_{ij1}=\ldots=\beta_{ijK}$ for all $i,j=1,2,3$.  Finally, by constructing confidence intervals around the parameter estimates, we implicitly consider the possibility that acoustic exposure may have effects only on specific contexts, or specific inter-state transitions. 

\subsubsection{Model fitting}

The likelihood of the model including both discrete-valued random effects and the sonar exposure covariate has exactly the same form as stated in Eq.\ (\ref{like2}), except that now each t.p.m.\ additionally varies over dives as it depends on the associated covariate value $x_{wd}$.

As the mixed HMM requires estimation of even more parameters than the model with the discrete random effects, fitting it entails technical and numerical challenges equal to or greater than those encountered with the discrete random effect model.  To obtain reliable model parameter estimates, we followed the same procedure outlined in Section \ref{fancyfit}.

\subsubsection{Results}

Table \ref{table-AICs} provides model selection results, including the models with common and context-specific acoustic exposures.  The best model as selected by AIC has $N$ = 3 states, $K$ = 4 behavioural contexts, and an effect of acoustic exposure that is common across all behavioural contexts.    

\begin{figure}[!htb]
\centering
\makebox{\includegraphics[width=\textwidth]{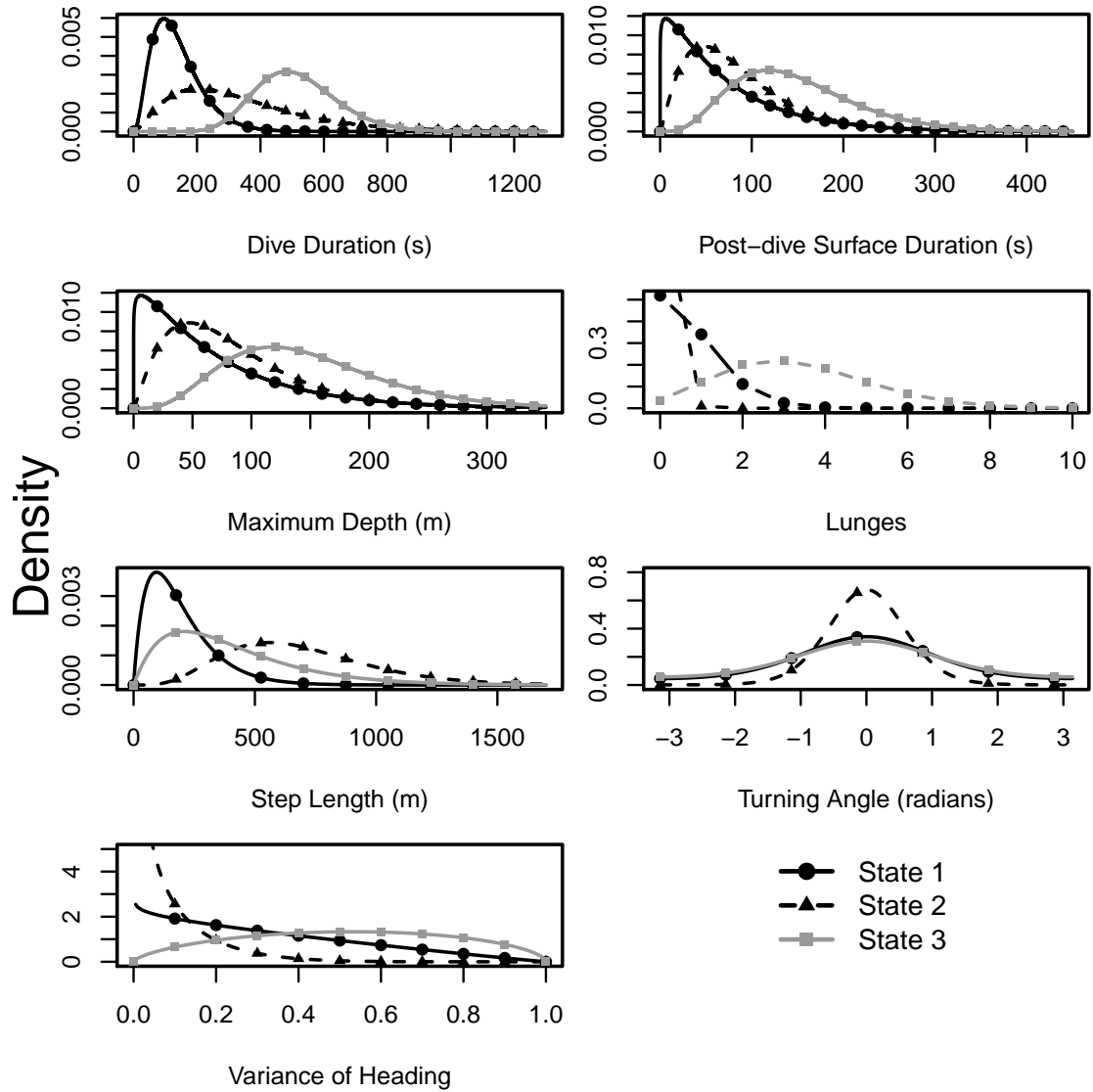}} \caption{\label{fig-state-dists2} Fitted model estimates of the state-dependent distributions for the 3-state HMM with 4 contexts and a common effect of acoustic disturbance.}
\end{figure}

Figure \ref{fig-state-dists2} shows the state-dependent distributions for the final (best) model.  State 1 includes the shortest, shallowest dives, with short step lengths and few lunges; it may correspond to shallow foraging behaviour.  State 2 features intermediate depth and duration, long step length, small turning angle, low heading variance, and no lunges; perhaps it can be labelled travelling behaviour.  State 3 has the deepest, longest dives, with intermediate step length, large variability of heading, and multiple lunges; it is consistent with deep foraging behaviour. These distributions are very similar to those of the initial model (with $N$ = 3 and $K$ = 1; Figure \ref{fig-state-dists}), confirming that the main differences between the initial and final models are in the transition probabilities, not the properties of the states themselves.

\begin{figure}[!htb]
\centering
\makebox{\includegraphics[width=\textwidth]{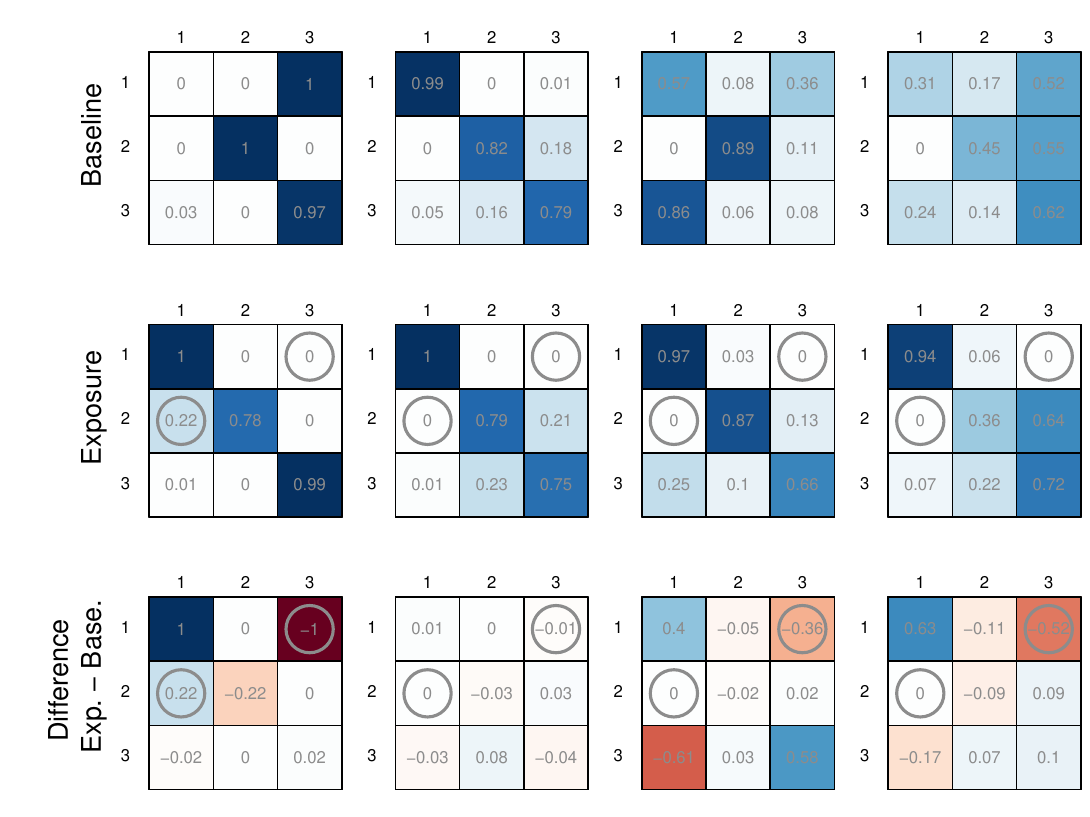}} \caption{Fitted model estimates of the transition probability matrices for the 3-state HMM with 4 contexts (one column per context).  For each context, the first row shows the  t.p.m. (in the absence of acoustic disturbance), the second row shows the t.p.m. with acoustic disturbance, and the third row shows the difference between the two.  The magnitudes of the transition probabilities are indicated by the fill colour, as well as the printed numeric values in each cell. For the t.p.m.s with acoustic disturbance, circles around the probabilities indicate entries corresponding to acoustic-disturbance parameters whose 95\% confidence intervals do not contain zero. }
\label{fig-best-tpms} 
\end{figure}

For the final, best model, the baseline t.p.m.\ is different for each of the four behavioural contexts, and each of the four matrices is affected in the same way by acoustic disturbance; the resulting eight baseline and exposure t.p.m.s are presented graphically in Figure \ref{fig-best-tpms}. Table \ref{table-deltas} presents the initial state distributions for each context.  Context 1 has very high persistence in states 2 and 3, and immediate transition from state 1 to 3 (if 1 is ever observed); context 2 has high persistence in all three states, but with occupancy of state 3 quite rare (cf.\ Table \ref{table-deltas}); context 3 has persistence in state 2 and some alternation between states 1-3; and context 4 has the highest probability of switching between states, with frequent transitions to state 3 from other states.

\begin{table}[!htb]
\caption{Initial state distributions $ \boldsymbol{\delta}^{(w)}$ for the best model, with $N$ = 3, $K$ = 4, and a common effect of acoustic disturbance.}
\vspace{0.5em}
\label{table-deltas}
\centering
\begin{tabular}{cccc}
  \hline
Context & State 1 & State 2 & State 3 \\ 
  \hline
  $k=1$ & 0.84 & 0.00 & 0.16 \\ 
  $k=2$ & 0.63 & 0.37 & 0.00 \\ 
  $k=3$ & 0.50 & 0.50 & 0.00 \\ 
  $k=4$ & 0.66 & 0.00 & 0.34 \\ 
   \hline
\end{tabular}
\end{table}

\begin{table}
\caption{Parameter estimates, along with 95\% CIs, for the best model as selected by AIC ($N$ = 3 states, $K$ = 4 behavioural contexts, and an effect of acoustic exposure that is common across all behavioural contexts).} \label{table-best-params}
\begin{center}
\begin{tabular}{lcccc}
\hline \hline \\[-2ex]
   \multicolumn{1}{l}{{Parameter}} &
   \multicolumn{1}{c}{{State}} &
   \multicolumn{1}{c}{{Lower Bound}} &
   \multicolumn{1}{c}{{Estimate}} &
   \multicolumn{1}{c}{{Upper Bound}} \\[0.5ex] \hline
   \\[-1.8ex]
Dive Duration ($\mu$)        & 1        & 131.496     & 138.817  & 146.546                \\
Dive Duration ($\sigma$)     & 1        &  71.406     &  77.784  &  84.731                \\
Dive Duration ($\mu$)        & 2        & 312.358     & 342.161  & 374.808                \\
Dive Duration ($\sigma$)     & 2        & 190.224     & 216.785  & 247.056                \\
Dive Duration ($\mu$)        & 3        & 502.146     & 514.913  & 528.004                \\
Dive Duration ($\sigma$)     & 3        & 119.634     & 129.102  & 139.320                \\
Surface Duration ($\mu$)     & 1        &  65.183     &  71.250  &  77.881                \\
Surface Duration ($\sigma$)  & 1        &  61.179     &  68.068  &  75.731                \\
Surface Duration ($\mu$)     & 2        &  76.394     &  84.765  &  94.054                \\
Surface Duration ($\sigma$)  & 2        &  48.202     &  56.487  &  66.197                \\
Surface Duration ($\mu$)     & 3        & 143.200     & 149.973  & 157.066                \\
Surface Duration ($\sigma$)  & 3        &  63.573     &  69.175  &  75.270                \\
Maximum Depth ($\mu$)        & 1        &  30.011     &  32.181  &  34.508                \\
Maximum Depth ($\sigma$)     & 1        &  21.242     &  23.344  &  25.654                \\
Maximum Depth ($\mu$)        & 2        &  58.600     &  67.662  &  78.124                \\
Maximum Depth ($\sigma$)     & 2        &  54.265     &  64.385  &  76.393                \\
Maximum Depth ($\mu$)        & 3        & 163.521     & 169.459  & 175.613                \\
Maximum Depth ($\sigma$)     & 3        &  56.329     &  60.859  &  65.755                \\
Number of Lunges ($\lambda$) & 1        &   0.492     &   0.656  &   0.651                \\
Number of Lunges ($\lambda$) & 2        &   0.001     &   0.010  &   0.095                \\
Number of Lunges ($\lambda$) & 3        &   3.064     &   3.312  &   3.431                \\
Step Length ($\mu$)          & 1        & 179.387     & 192.978  & 207.599                \\
Step Length ($\sigma$)       & 1        & 124.894     & 137.948  & 152.366                \\
Step Length ($\mu$)          & 2        & 648.927     & 693.370  & 740.857                \\
Step Length ($\sigma$)       & 2        & 269.216     & 304.052  & 343.395                \\
Step Length ($\mu$)          & 3        & 379.235     & 408.935  & 440.960                \\
Step Length ($\sigma$)       & 3        & 258.701     & 286.763  & 317.868                \\
Turning Angle ($\kappa$)     & 1        &   0.850     &   1.007  &   1.193                \\
Turning Angle ($\kappa$)     & 2        &   2.611     &   3.175  &   3.862                \\
Turning Angle ($\kappa$)     & 3        &   0.681     &   0.841  &   1.038                \\
Heading Variance ($a$)       & 1        &   0.826     &   0.940  &   1.070                \\ 
Heading Variance ($b$)       & 1        &   1.774     &   2.046  &   2.358                \\ 
Heading Variance ($a$)       & 2        &   0.435     &   0.518  &   0.617                \\ 
Heading Variance ($b$)       & 2        &   4.889     &   6.552  &   8.779                \\ 
Heading Variance ($a$)       & 3        &   1.416     &   1.656  &   1.860                \\
Heading Variance ($b$)       & 3        &   1.342     &   1.578  &   1.827                \\   
Mass, rand.\ eff.\ comp.\ 1 ($\pi_1$) &   &   0.189     &   0.333  &   0.494          \\ 
Mass, rand.\ eff.\ comp.\ 2 ($\pi_2$) &   &   0.280     &   0.449  &   0.607          \\ 
Mass, rand.\ eff.\ comp.\ 3 ($\pi_3$) &   &   0.013     &   0.054  &   0.187          \\ 
Mass, rand.\ eff.\ comp.\ 4 ($\pi_4$) &   &   0.070     &   0.163  &   0.320          \\ 
Acoustic Exposure ($\beta_{12\cdot}$) &                 &  -2.555     &  -0.997  &   0.561                \\  
Acoustic Exposure ($\beta_{13\cdot}$) &                 &  $-\infty$    & -27.071  &  -6.855                \\  
Acoustic Exposure ($\beta_{21\cdot}$) &                 &  15.022     &  17.008  &  18.993                \\  
Acoustic Exposure ($\beta_{23\cdot}$) &                 &  -0.829     &   0.147  &   1.123                \\  
Acoustic Exposure ($\beta_{31\cdot}$) &                 &  -3.215     &  -1.241  &   0.733                \\  
Acoustic Exposure ($\beta_{32\cdot}$) &                 &  -0.507     &   0.390  &   1.287   \\ \hline                
\end{tabular}
\end{center}
\end{table}

Table \ref{table-best-params} presents the maximum likelihood estimates of the parameters related to the state-dependent process, of the weights associated with the discrete random effects, and of the parameters describing the effect of sonar exposure. The table also includes 95\% confidence intervals for the parameter estimates.  These confidence intervals were obtained from the observed Fisher information, except for the interval provided for the parameter $\beta_{13\cdot}$, which was obtained using profile likelihood since the observed Fisher information indicated a standard error of zero for this estimator.  Among the acoustic exposure parameters ($\beta$s), only $\beta_{13\cdot}$ and $\beta_{21\cdot}$ have 95\% confidence intervals that do not contain zero.  The corresponding parameter estimates indicate that during acoustic exposure, the rate of already-rare transitions from state 1 to state 3 falls to zero ($\beta_{13\cdot}$ = -27.1), balanced by an increased probability of remaining in state 1.  There is also an increase in the rate of transitions from state 2 to state 1 ($\beta_{21\cdot}$), at the expense of a decreased persistence in state 2. 

For each whale, we decode the states locally (i.e.\ for each data point, we calculate the most likely state, under the fitted model and given the observations), as follows:
\begin{align*}
\Pr & \bigl( S_{wd} = j \mid \mathbf{X}_{w\cdot\cdot}, \mathbf{z}_{w\cdot}\bigr) \\ & = \sum_{k=1}^K \Pr \bigl( S_{wd} = j , \xi^{(w)}=k \mid \mathbf{X}_{w\cdot\cdot} , \mathbf{z}_{w\cdot} \bigr) \\
& = \sum_{k=1}^K \Pr \bigl( S_{wd} = j \mid \xi^{(w)}=k, \mathbf{X}_{w\cdot\cdot} , \mathbf{z}_{w\cdot} \bigr) \Pr \bigl( \xi^{(w)}=k \mid \mathbf{X}_{w\cdot\cdot} , \mathbf{z}_{w\cdot} \bigr) \\
& = \sum_{k=1}^K \Pr \bigl( S_{wd} = j \mid \xi^{(w)}=k, \mathbf{X}_{w\cdot\cdot} , \mathbf{z}_{w\cdot} \bigr) \frac{\mathcal{L}_k \pi_k}{\mathcal{L}}. \\
\end{align*}
Bayes' theorem is applied in the last step. The calculation of the first term simply corresponds to the local decoding of a basic HMM with no random effects and is performed as described in \citet{Zucchini2009} (Section 5.3.1). All remaining terms are evaluated as previously described. These decoded states are included in Figures \ref{fig-data1}-\ref{fig-data2}, and corresponding figures for all whales analysed are included in the web-based supplementary materials.

\section{Discussion}\label{discuss}
Our analysis of the blue whale data demonstrates the successful application of HMMs to multivariate animal behaviour data.  While multivariate HMMs and other hidden state models are already in use for ecological data, most applications involve fitting models to animal movement data, such as animal tracks in two or three dimensions (for example, see \cite{Langrock2012, McKellar2014, Langrock2014, McClintock2012, Patterson2008, Johnson2008, Morales2004}).  Examples of using HMMs to classify and characterise broader animal behaviour states are much less prevalent (but see \cite{Bagniewska2013, Isojunno2015}).  Here, we demonstrate the power and flexibility of this approach.  By applying HMMs to free-ranging marine mammals in a range of behavioural states and sound exposure regimes, we illustrate how they can incorporate random effects to account for individual or contextual differences, and also quantify the effects of environmental covariates (here, sound exposure) in modulating behavioural transitions.  An additional strength of this modelling approach, particularly for biological data sets, is its ability to simply and effectively deal with partially observed multivariate data without omitting any observed data points.

Our best model for the blue whale data incorporates a discrete random effect to account for differences between observed whales, whether these are due to genuine individual variation or the effects of unobserved contextual variables.  Large inter-individual variability is a prominent, obvious feature of our dataset and many other ecological datasets.  In fact, the magnitude of these differences is often equal to or greater than the changes induced by experimental manipulation, so accounting for them can dramatically improve model fit, as evidenced by the very large improvement in AIC score between models without and with the random effect.  More importantly, failing to account for individual differences (or contexts, as we call them here) may make it difficult to detect relatively small-magnitude behaviour modulations caused by other factors (such as acoustic disturbance). In particular, not accounting for individual heterogeneity can result in biases and invalid standard errors of estimators that describe an effect of interest. While it is easy to argue that random effects to account for individual variation are a key component of animal behaviour models, fitting traditional continuous random effects can be computationally challenging, especially for models like HMMs that are already complex and parameter-rich.  The discrete random effect formulation used in this study makes inclusion of between-subject variability, which is key for improving model fit, practically feasible and reasonably quick, within a simple maximum-likelihood estimation framework.    

In addition to accounting for inter-subject differences, we illustrate here a simple formulation allowing the inclusion of a covariate affecting the t.p.m., which could easily be generalised to multiple covariates if the data allow.  In the blue whale case, previous research suggested that the most likely effect of acoustic disturbance would be on rates of transition between behavioural states \citep{Goldbogen2013d}. However, one could also consider including covariates affecting one or more of the state-dependent distributions, if appropriate for the data. In the case of acoustic disturbance in particular, it would be of considerable biological interest to consider whether response intensity scales with stimulus type (simulated or real MFAS, or PRN) or with the received level of the sound stimulus at the whale.  In fact, while many regulations assume that responsiveness depends on exposure level, considerable and evolving data support the conclusion that myriad contextual factors related to the exposure configuration and internal behavioural features of the subject can strongly affect response probability \citep{Ellison2012}. A wealth of studies indicate differential responses to different sound types within the same species (for example, \cite{Miller2012, Goldbogen2013d, Nowacek2004}, and reviews \cite{Southall2007, DeRuiter2010}), although in some cases sound frequency and type have not been confirmed as important predictors of response \citep{Antunes2014}.  Potential effects of signal type and received level could be easily accommodated in our model formulation, essentially resulting in a multiple multinomial logistic regression of the transition probabilities on the stimulus type and intensity.  Unfortunately, sample-size but also computational limitations prevent us from implementing these models and fitting them to the blue whale dataset. This circumstance highlights one limitation of hidden state models like our mixed multivariate HMM: they are parameter-rich and can be challenging to fit to data, such that computational considerations and sample size often limit the complexity of candidate models. It is therefore interesting, and perhaps surprising, that we see a relatively consistent behavioural response in our data across several stimulus types (as further detailed below).  One reason may be the relatively limited range of exposure levels; none of the CEEs resulted in received levels exceeding 157 dB re. 1 $\mu$Pa rms (computed as detailed in \citet{DeRuiter2013a}), so relatively little information is lost in considering the acoustic disturbance as ``on" or ``off" and disregarding the measured intensity. 

In this study of blue whale behavioural responses to acoustic disturbance, it is difficult to draw firm conclusions about the biological significance of the changes observed during acoustic exposure: the difference in AIC scores between the model with and without the exposure covariate is only about 5.1.  While the model with an effect of acoustic disturbance has considerably more support from the data than the model without, the improvement in model fit resulting from the exposure covariate is modest compared to that of the random effect.  This modest improvement could result from the limited number of CEE dives observed.  It might also indicate that the disturbance effect has a relatively small magnitude, and/or affects only some of the individuals, and/or affects only certain behaviour states, as evidenced by the confidence intervals presented in Table \ref{table-best-params}. These results are again consistent with those of \citet{Goldbogen2013d}.

The mixed HMM fitted here allows us not only to state that the acoustic disturbance had an effect, but also to characterise that effect quantitatively.  The types of changes that occur, according to the common acoustic exposure effect model, are consistent with previous work.  We find that whales are less likely to initiate deep foraging behaviour during acoustic disturbance, in accordance with \citet{Goldbogen2013d}, who analysed a subset of the data presented here; beaked whales \citep{DeRuiter2013a, Stimpert2014, Miller2015, Tyack2011} and sperm whales \citet{Isojunno2015} have also been shown to avoid deep-feeding behaviour during disturbance, although in each case more research is needed to understand whether this response helps whales conserve energy, avoid detection, or something else. We also found that the rate of transition from state 2 (directed travel) to state 1 (shallow dives, with no or few feeding lunges) was elevated during exposure, but only in behavioural context 1.  (In all other contexts, the baseline probability of transition from state 2 to 1 was so low that the ``elevated" probability during exposure was still effectively zero.)  This type of change might relate to increased vigilance or variability of behaviour during lower-level acoustic disturbance related to consideration of an appropriate course of action, as seen in some beaked whales \citep{DeRuiter2013a, Miller2015}.  However, since this observation relates to only one context and thus a relatively small number of whales and observations, more research would clearly be needed to confirm these results and their interpretation. 

Another important consideration for the interpretation of the model results is the grouping of dives with and without lunges together in state 1, as it relates to our observations of the natural history of these whales. By selecting a three-state model to reasonably group dives into categories, the model groups no-, single-, and several-lunge dives within a single shallow-diving state, indicating that these dives are quite similar (aside from number of lunges), or at least that the differences between them are less than those between shallow, travelling, and deep feeding dives. While this model categorization may differ slightly from a priori distinctions between shallow feeding, deep feeding, and non-feeding \citep[as in ][]{Goldbogen2013d}, we believe this grouping is still generally consistent with the behavioural state of shallow-diving and feeding behaviour in blue whales. Recent data, for instance, indicates that there may be significantly greater energetic benefit (as well as cost) in deep versus shallow foraging \citep{Hazen2015}.  

The results of this study are consistent with previous observations that behavioural responsiveness to acoustic disturbance is highly context-specific: animals modulate the intensity of their response depending not only on the intensity of the sound they experience, but also on environmental and behavioural \citep{Ellison2012}.  This study illustrates that even if the type of response is consistent across behavioural contexts, the apparent intensity of the response displayed will vary depending on the baseline behaviour (Figure \ref{fig-best-tpms}). The detection and further characterisation of these behavioural responses is of strong biological and management interest, since it can contribute to our understanding and mitigation of any potential negative effects of anthropogenic noise.  Since the behaviour changes observed relate to alterations of foraging behaviour, they have particular potential to affect animals' health and fitness and perhaps lead to population consequences.  Mixed HMMs should prove a flexible, powerful tool for the characterisation of animal behavioural states, including behavioural responses to acoustic disturbance or other environmental stimuli.

\section*{Acknowledgments} We thank all the participants in field work for the SOCAL-BRS project between 2010-2014 (identified in greater detail at: www.socal-brs.org), without whom these valuable data would not have been collected.  This research was supported by the U.S. Office of Naval Research grant N00014-12-1-0204, under the project entitled Multi-study Ocean acoustics Human effects Analysis (MOCHA).  We thank the MOCHA working group, especially L. Thomas, C. Harris and D. Sadykova, for input and feedback at various stages of analysis development. Funding was also provided to the SOCAL project by the U.S. Navy Chief of Naval Operations, Environmental Readiness Program through the Living Marine Resources Program, the U.S. Office of Naval Research Marine Mammal Research Program, and the National Research Council.  Experiments were conducted in compliance with the terms of NMFS permit no. 14534 (B. Southall, principal investigator).   

\renewcommand\refname{References}
\makeatletter
\renewcommand\@biblabel[1]{}

\addcontentsline{toc}{section}{References} 
\bibliographystyle{apalike}
\bibliography{library}

 \end{spacing}
\end{document}